\documentclass{aa}
\usepackage{graphicx}
\usepackage[varg]{txfonts}
\usepackage{subfigure}
\usepackage{url}
\usepackage[citecolor=blue,colorlinks=true]{hyperref}
\usepackage{ulem}

\begin{document} 

\title{Upper limits on the water vapour content \\ of the $\beta$~Pictoris debris disk\thanks{{\it Herschel} is an ESA space observatory with science instruments provided by European-led Principal Investigator consortia and with important participation from NASA.}}

\author{
M. Cavallius\inst{\ref{StockholmUniversity}}
\and
G. Cataldi\inst{\ref{StockholmUniversity},\ref{Budapest},\ref{Subaru},\ref{Toronto}}\thanks{International Research Fellow of Japan Society for the Promotion of Science (Postdoctoral Fellowships for Research in Japan (Standard)).}
\and
A. Brandeker\inst{\ref{StockholmUniversity}}
\and
G. Olofsson\inst{\ref{StockholmUniversity}}
\and
B. Larsson\inst{\ref{StockholmUniversity}}
\and
R. Liseau\inst{\ref{Onsala}}
}

\institute{
Department of Astronomy, AlbaNova University Center, Stockholm University, 106 91 Stockholm, Sweden, \email{[maria.cavallius; alexis; olofsson; bem]@astro.su.se} \label{StockholmUniversity}
\and
Konkoly Observatory, Research Centre for Astronomy and Earth Sciences, Hungarian Academy of Sciences, Konkoly-Thege Mikl\'os \'ut 15--17, 1121 Budapest, Hungary\label{Budapest}
\and
Subaru Telescope, National Astronomical Observatory of Japan, 650 North A'ohoku Place, Hilo, HI 96720, USA \label{Subaru} 
\and
Department of Astronomy and Astrophysics, University of Toronto, Toronto, ON M5S 3H4, Canada, \email{cataldi.gia@gmail.com}\label{Toronto}
\and
Department of Space, Earth and Environment, Chalmers University of Technology, Onsala Space Observatory, 439 92 Onsala, Sweden, \email{rene.liseau@chalmers.se} \label{Onsala}
}

\date{Received April 10, 2019; accepted June 23, 2019}

\abstract
   {The debris disk surrounding $\beta$~Pictoris has been observed with ALMA to contain a belt of CO gas with a distinct peak at $\sim$85\,au. This CO clump is thought to be the result of a region of enhanced density of solids that collide and release CO through vaporisation. The parent bodies are thought to be comparable to solar system comets, in which CO is trapped inside a water ice matrix.}
   {Since H$_2$O should be released along with CO, we aim to put an upper limit on the H$_2$O gas mass in the disk of $\beta$~Pictoris.}
   {We use archival data from the Heterodyne Instrument for the Far-Infrared (HIFI) aboard the Herschel Space Observatory to study the ortho-H$_2$O 1$_{10}$--1$_{01}$ emission line. The line is undetected. Using a python implementation of the radiative transfer code \texttt{RADEX}, we convert upper limits on the line flux to H$_2$O gas masses. The resulting lower limits on the CO/H$_2$O mass ratio are compared to the composition of solar system comets.}
   {Depending on the assumed gas spatial distribution, we find a 95\% upper limit on the ortho-H$_2$O
   line flux of $7.5 \times 10^{-20}$\,W\,m$^{-2}$ or $1.2 \times 10^{-19}$\,W\,m$^{-2}$.
   These translate into an upper limit on the H$_2$O mass of $7.4 \times 10^{16}$--$1.1 \times 10^{18}$\,kg depending on both the electron density and gas kinetic temperature. The range of derived gas-phase CO/H$_2$O ratios is
   marginally consistent with low-ratio solar system comets.}
   {}
   

\maketitle


\section{Introduction}
Circumstellar debris disks are made up of planetesimals and dust and can be thought of as the remnants of planet formation. The small dust grains in debris disks are continuously removed by radiation pressure on time-scales that are short compared to the age of the system. Thus, the dust must be secondary and is replenished through collisions between larger bodies \citep[e.g.][]{Wyatt_review,Wyatt2018}. Therefore, there exists a direct connection between the observed dust and the building blocks of planets. The study of debris disks is part of our efforts to understand the formation and evolution of planetary systems.

For a small subset of debris disks, gas has been detected in addition to the dust. Thanks to sensitive observations of CO emission with the Atacama Large Millimeter/submillimeter Array (ALMA), the set of known gaseous debris disks was recently expanded significantly \citep[e.g.][]{Lieman-Sifry2016,Moor2017}. Most of the $\sim$20 gaseous debris disks we currently know are hosted by young A-type stars. In general, the origin of this gas remains unknown. It could be primordial (i.e.\ leftover from the protoplanetary phase) or, like the dust, secondary (i.e.\ released from volatile-rich, icy bodies). For a few systems (e.g.\ $\beta$~Pictoris, see below), there is evidence for a secondary origin of the gas. Studying this secondary gas allows us to constrain the composition of the parent bodies and compare it to solar system comets \citep{Xie2013, Kral2016,Marino2016,Matra2017,Matra2017_Fomalhaut,Matra2018}.

One example of a gaseous debris disk is found around the A-type star $\beta$~Pictoris. This system has been the subject of numerous studies over the last three decades, with the IRAS discovery of the infrared (IR) excess of its debris disk dating back to 1984 \citep{Smith1984}. It is nearby \citep[19.75\,pc,][]{GaiaDR2}, young \citep[23 $\pm$ 3\,Myr,][]{Mamajek2014} and harbours a giant planet at $\sim$10\,au, discovered by direct imaging \citep{Lagrange2008}. The belt of parent bodies of the disk is located at around 100\,au \citep{Dent2014}, and is edge-on, making it an ideal target for absorption spectroscopy. The system has served as a laboratory to study various processes in a newly formed planetary system.

Several atomic metal lines have been detected in the gas phase of the $\beta$~Pic disk. The lines appear in both absorption against the star and in emission. Besides species such as Fe, Na and Ca$^+$ \citep{Brandeker2004}, this includes oxygen \citep{Brandeker2016} as well as both neutral and ionised carbon \citep{Roberge2006,Cataldi2014,Higuchi2017,Cataldi2018}. CO is the only molecule detected so far \citep[e.g.][]{Liseau1998,Dent2014,Matra2018}. In addition, \citet{Wilson_etal_2018} recently detected N in absorption and \citet{Wilson2017} H in close vicinity to the star (i.e., not in the main gas/dust disk). Elemental abundances differ significantly from solar values. In particular, there is an overabundance of C \citep{Roberge2006,Cataldi2014}, and perhaps O \citep{Brandeker2016}, relative to other elements.

Different pieces of evidence suggest that the gas in the $\beta$~Pic disk is secondary. First, the spatial distributions of gas and dust are similar \citep[e.g.][]{Brandeker2004,Dent2014}, consistent with a common origin. Second, the dynamical lifetime of the gas is short compared to the age of the system \citep{Fernandez2006}. Additional evidence comes from CO. \citet{Matra2017} showed that there is not enough H$_2$ in the disk to shield CO from photodissociation over the age of the system. The photodissociation lifetime of CO is thus only\,$\sim$50 years \citep{Cataldi2018}, implying a replenishment mechanism. Therefore, the CO is believed to be produced from volatile-rich cometary bodies. At least part of the observed C and O must therefore come from CO photodissociation.

The CO emission from the $\beta$~Pic disk was first resolved by ALMA \citep{Dent2014}. The data revealed a surprising asymmetry: a CO clump on the south-west (SW) side of the disk at a radial distance of\,$\sim$85\,au. The clump is interpreted as a site of enhanced CO gas production caused by an enhanced collision rate among volatile-rich bodies. The 
suggested mechanisms for the enhanced collision rate are a yet unseen, outward-migrating planet capturing comets in a resonance trap \citep{Matra2017}; a previous giant collision \citep{Jackson2014}; or an eccentric disk produced from a tidal disruption event \citep{Cataldi2018}.

Solar system comets consist mainly of amorphous water ice in a matrix structure, wherein volatiles like CO become trapped \citep{Whipple1950}. If the CO seen in the $\beta$~Pic disk is indeed produced by colliding comet-like bodies, we would expect water vapour to be produced as well, e.g. by photosputtering of grains formed in a collisional cascade, see Sect. \ref{sec:discussion}. Based on the excitation conditions of \ion{O}{i}, \citet{Kral2016} indirectly constrained the H$_2$O/CO ratio of the $\beta$~Pic comets. They find a lower ratio than what is observed for solar system comets, suggesting that the comets in the $\beta$~Pic disk are depleted in water. However, their assumed spatial distribution of the gas (accretion disk) turned out to be inconsistent with new observations \citep{Cataldi2018}. It is unclear how this impacts their estimated H$_2$O/CO ratio.

\citet{Matra2017,Matra2017_Fomalhaut} and \citet{Marino2016} found that the CO+CO$_2$ mass abundances of the comets in the $\beta$~Pic, Fomalhaut and HD~181327 disks are consistent with abundances measured in solar system comets. Using data from the Submillimeter Array (SMA), \citet{Matra2018} showed that the comets in the $\beta$~Pic disk have a HCN/(CO+CO$_2$) ratio of outgassing rates that is consistent with, but at the low end of, what is observed for solar system comets. This might be caused by the outgassing mechanism, or it might indicate a true depletion of HCN with respect to solar system comets.

One motivation to study the water content of extrasolar planetesimals comes from an astrobiological perspective. Water is required for life as we know it. Most of the water on Earth may have been delivered by asteroids \citep[e.g.][]{Alexander2012}, and similar processes might occur in extrasolar systems. In this work, we aim to further investigate the water content of the bodies in the $\beta$~Pic disk with a more direct approach compared to the \citet{Kral2016} study. We use an archival non-detection from \textit{Herschel}/HIFI of the rotational 1$_{10}$--1$_{01}$ transition of o-H$_2$O. Our paper is organised as follows: In Sect.~\ref{sec:observations}, we describe our observations. We derive upper limits on the o-H$_2$O 1$_{10}$--1$_{01}$ flux and the H$_2$O mass in Sect.~\ref{sec:analysis}, and report them in Sect.~\ref{sec:results}. In Sect.~\ref{sec:discussion}, we discuss what our measurements tell us about the water content of the bodies in the $\beta$~Pic disk. We summarise our
results in Sect.~\ref{sec:conclusions}.


\section{Observations and data reduction}\label{sec:observations}
$\beta$~Pic was observed on July 19th 2010 (ID 1342200907) with the Heterodyne Instrument for the Far-Infrared \citep[HIFI,][]{HIFI} aboard \textit{Herschel} \citep{Pilbratt2010}, as a Guaranteed Time observation. The Single Point mode of the High Resolution Spectrometer was used, and the observation time was roughly 58\,min. The lower sideband of band 1b was chosen in order to observe the o-H$_2$O 1$_{10}$--1$_{01}$ line, which has a rest frame frequency of 556.936\,GHz \citep{NIST}. The spectral resolution was 125 kHz. The FWHM beam size was 37\arcsec. The clump of CO is at a separation of 4.3\arcsec\ from the center of the beam, meaning the beam sensitivity at its location was 96\,\%. As the bulk of the CO as determined by \citet{Dent2014} is within this angular distance, we do not perform any beam sensitivity correction in the data analysis.

The data were first processed in the Herschel Interactive Processing Environment \citep[HIPE version 13.0.0,][]{HIPE}. The antenna temperature was converted to a flux density by assuming a point source. Standing waves were removed and a baseline correction was carried out. The four sub-bands were stitched together and since the Frequency Shift mode was used, the spectra were folded. The H and V polarisations were then averaged over. The spacing of data points was 0.12\,MHz, but to ensure there were no remaining standing waves to correlate adjacent flux points, we calculated the auto-correlation in the background region (see below) using numpy's built-in \texttt{correlate} function. The FWHM of the best-fit gaussian of the correlation function was then taken to be the correlation length, and was 0.3631\,MHz. The spectrum was then rebinned to 0.5\,MHz to ensure the bins were independent of one another. Figure \ref{fig:bestfit} shows the reduced spectrum in a velocity scale relative to the $\beta$~Pic system, where the star has a heliocentric velocity of 20.5\,km\,s$^{-1}$ \citep{Brandeker2011}.


\section{Analysis}\label{sec:analysis}
\subsection{Upper limit on the \texorpdfstring{o-H$_2$O 1$_{10}$--1$_{01}$}{o-H2O 110-101} flux}\label{subsec:flux_upper_limit}

We assume that H$_2$O and CO are released at the same locations in the disk and subsequently photodissociated on timescales short compared to the orbital timescale. At 85\,au, the CO lifetime is $\sim$50\,yr. The lifetime of H$_2$O is even shorter: using the stellar spectrum supplemented with UV data described in \citet{Cataldi2018} and photodissociation cross sections from the Leiden Observatory database of the `photodissociation and photoionisation of astrophysically relevant molecules'\footnote{\url{http://home.strw.leidenuniv.nl/~ewine/photo/}} \citep{Mordaunt1994,vanHarrevelt2008,Heays2017}, we calculate a lifetime of 3.5\,d under the assumption that water is not shielded from dissociating radiation. This assumption is justified since neither CO nor C can shield water, as their cross sections for dissociation respectively ionisation do not substantially overlap in wavelength with the cross section for water dissociation. Furthermore, we verified that water self-shielding is irrelevant for the column densities considered here. This implies that, while the CO may have some time to orbitally evolve before being dissociated (the orbital period at 85\,au is $\sim$600\,yr), any water vapour must be close to its point of release. 

In one of the proposed CO production scenarios, the production region is caused by debris in eccentric
orbits converging at a fixed region in space corresponding to the south-western CO-`clump'. Orbital
motion then moves the gas out of this region forming a tail \citep{Dent2014}. Due to the shorter lifetime
of H$_2$O, no water vapour tail would be formed and all H$_2$O is expected to be restricted to the clump
region.
In the other proposed scenario, the CO production region is due to a mean motion resonance with an inner
planet, in which case the production region is more spread out \citep{Matra2017}. To accommodate both scenarios, we test for two extremes: first with the assumption that the bulk of the gas production takes place in the clump region, and second using the assumption that the CO distribution reflects the production region, in which case the H$_2$O
would share the spatial distribution of the CO.

\begin{figure}
\centering
\resizebox{\hsize}{!}
{\includegraphics{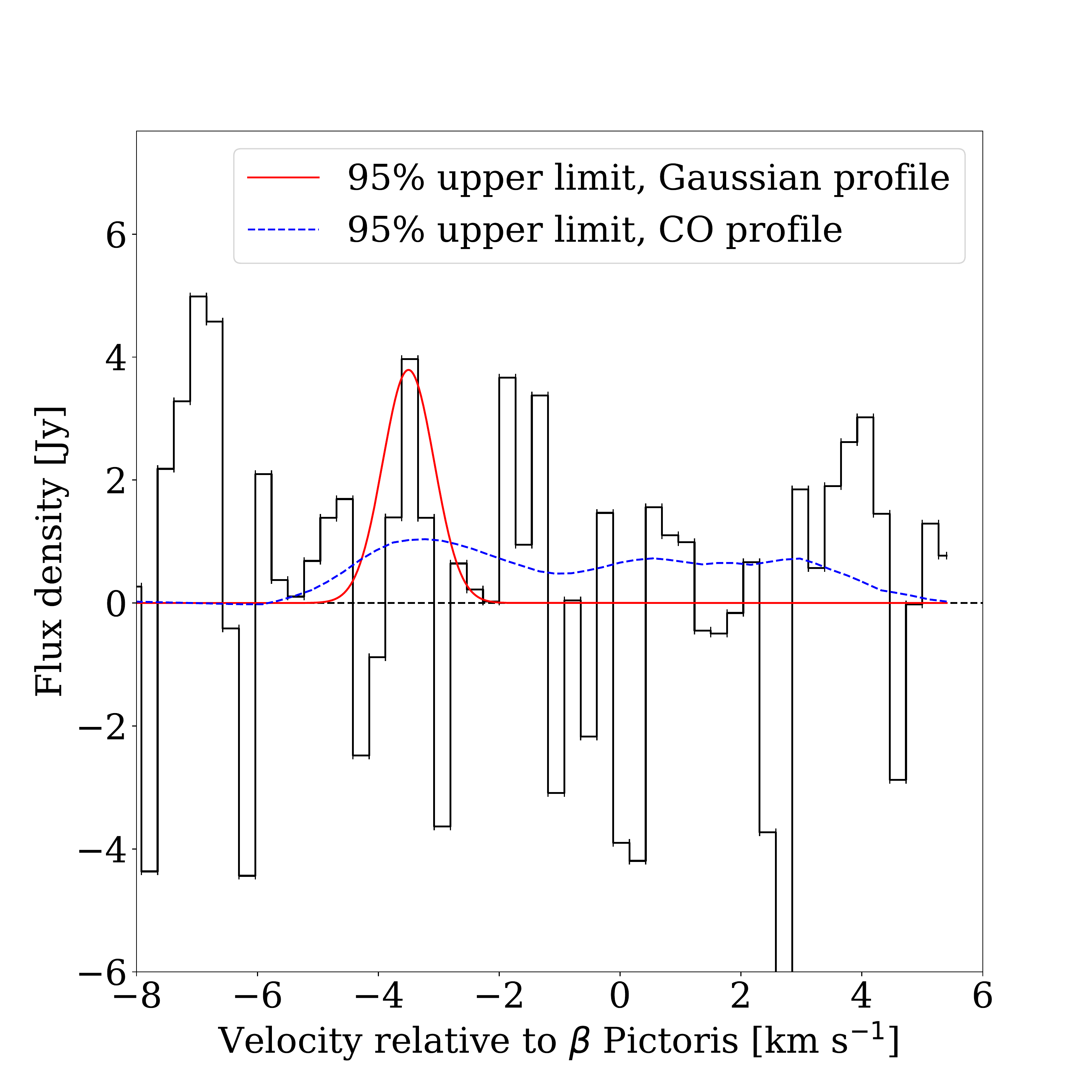}}
\caption{95\% upper limit models superimposed on the observed spectrum. The red, solid line shows the 
expected profile coming from H$_2$O restricted to the SW clump. The blue, dashed line shows the profile
from the model assuming the same spatial distribution for H$_2$O as for CO (see Sect. \ref{sec:discussion}).}
\label{fig:bestfit}
\end{figure}

The spectrum does not contain any clear emission line.
To estimate a reliable upper limit of the line flux, a Monte
Carlo approach is used. $10^6$ noise realisations are made by scaling a vector of random numbers
drawn from a standard normal distribution by the error $\sigma_\mathrm{i}$ of the flux density,
estimated as the standard deviation of the background flux density, where the background region is taken to be
\mbox{v > -0.56\,km\,s$^{-1}$} or v \mbox{$< -6.2$\,km\,s$^{-1}$}. Each noise realisation is then separately
added to the original spectrum to produce $10^6$ spectra with different noise realisations. 
A line profile is then fit by a $\chi^2$ minimisation in each alternate spectrum to obtain a best-fit flux. As line profiles we use the ones corresponding to the two scenarios (Fig.~\ref{fig:bestfit}): in the case of H$_2$O located in a clump we assume a Gaussian blue-shifted to -3.5\,km\,s$^{-1}$ with a full width at half maximum (FWHM) matching that of the CO, namely 1\,km\,s$^{-1}$ \citep{Dent2014}. For the case where H$_2$O shares the spatial distribution of CO, we use the same spectral profile as found for CO~3--2 \citep{Matra2017}. Since the shapes of the profiles are kept fixed, the only free parameter is the integrated flux, which is allowed to take on negative values. In this way a Gaussian distribution of $10^6$ values of the integrated line flux for each case is computed, from which the 95\% upper limits are derived.

\subsection{Upper limit on the \texorpdfstring{H$_2$O}{H2O} mass}
To convert the upper limits on the line flux into upper limits on the H$_2$O mass, we use \texttt{pythonradex}\footnote{\url{https://github.com/gica3618/pythonradex}}, a python implementation of the \texttt{RADEX} code \citep{Radex}. The program uses an escape probability formalism and an Accelerated Lambda Iteration (ALI) scheme to solve the radiative transfer in the non-Local Thermal Equilibrium (non-LTE) regime, given the total gas mass, and the densities and kinetic temperature of the colliders. In this way we are able to calculate the expected line flux for a given ortho-H$_2$O gas mass. To convert to total gas mass, an H$_2$O ortho:para ratio of 3:1 is assumed.

The code assumes a uniform, spherical geometry (we choose a radius of 30\,au) and that the line profile has a boxcar shape (we choose a width of 1.5\,km\,s$^{-1}$). Our model is thus not fully self-consistent, since we used different line profiles to compute upper limits on the line flux (Fig.~\ref{fig:bestfit}). 
The assumption of a spherical rather than a flattened, edge-on distribution may be relevant if the emitting region is optically thick. We study this possibility in Sect.~\ref{sec:discussion}, resulting in Fig.~\ref{fig:clumpmass}.

We assume that electrons are the only relevant colliders with H$_2$O \citep[e.g.][]{Kral2016}. While a certain
amount of hydrogen should also be present (e.g.\ from H$_2$O photodissociation), the collisional excitation
rates with H are several orders of magnitude lower compared to electrons (A.\ Faure, private communication).
We use excitation rates from the LAMDA database\footnote{\url{http://home.strw.leidenuniv.nl/~moldata/}}
\citep{Schoier2005} published by \citet{Faure2004}, supplemented at kinetic temperatures below 200\,K with
unpublished data (A.\ Faure, private communication). In the low-temperature regime the data consist of
excitation rates of transitions between the 14 lowest-lying levels of o-H$_2$O, while the high-temperature
regime includes 397 additional levels for a total of 411. The o-H$_2$O 1$_{10}$--1$_{01}$ emission line
corresponds to the transition between the two lowest levels. 
We confirm that no levels other than the two lowest are significantly populated,
which could otherwise lead to fluorescence effects. 
In any case, if fluorescence was important, neglecting it would result in an underestimate of 
the line flux from the transition and therefore an overestimated upper limit on the mass.

In addition to collisional excitation from electrons we also include a background radiation field consisting of the cosmic microwave background, stellar photons directly from $\beta$~Pic, and a model of the dust radiation field. The latter is as shown in Fig.~B1 of \citet{Kral2017} and represents the dust radiation field in the midplane of the disk at the position angle of the clump (L.\ Matr\'{a}, private communication). The e$^-$--H$_2$O collisions are by far the most important excitation mechanism, hence including the background field does not change the results significantly.

\citet{Matra2017} studied the CO J=3--2/J=2--1 line ratio and found the gas kinetic temperature to be between 40 and a few hundred\,K, and the electron density in the disk to be around $10^2$--$10^3$\,cm$^{-3}$. The temperature and electron densities are degenerate with their estimates depending on the assumed mean molecular weight of the gas. We therefore consider four different temperatures (50, 100, 200 and 400\,K) and three different electron densities (100, 300 and 900\,cm$^{-3}$). In addition, we also perform a calculation assuming LTE. For a given electron density and temperature, we then compute the o-H$_2$O 1$_{10}$--1$_{01}$ flux as a function of mass. 


\section{Results}\label{sec:results}

The resulting upper limit on the o-H$_2$O 1$_{10}$--1$_{01}$ flux depends on the assumed line profile. 
With all water vapour constrained to the clump, the narrower spectral line results in a 
95\% upper limit of $7.5 \times 10^{-20}$\,W\,m$^{-2}$, while the corresponding limit for the 
broader spectral line coming from the spread-out scenario is $1.2 \times 10^{-19}$\,W\,m$^{-2}$.
Figure~\ref{fig:bestfit} shows a plot of the spectrum after the initial processing in HIPE with 
the spectral lines from both scenarios plotted at their 95\% upper limit flux.

\begin{figure*}
	\centering
	\resizebox{\hsize}{!}
	{\includegraphics{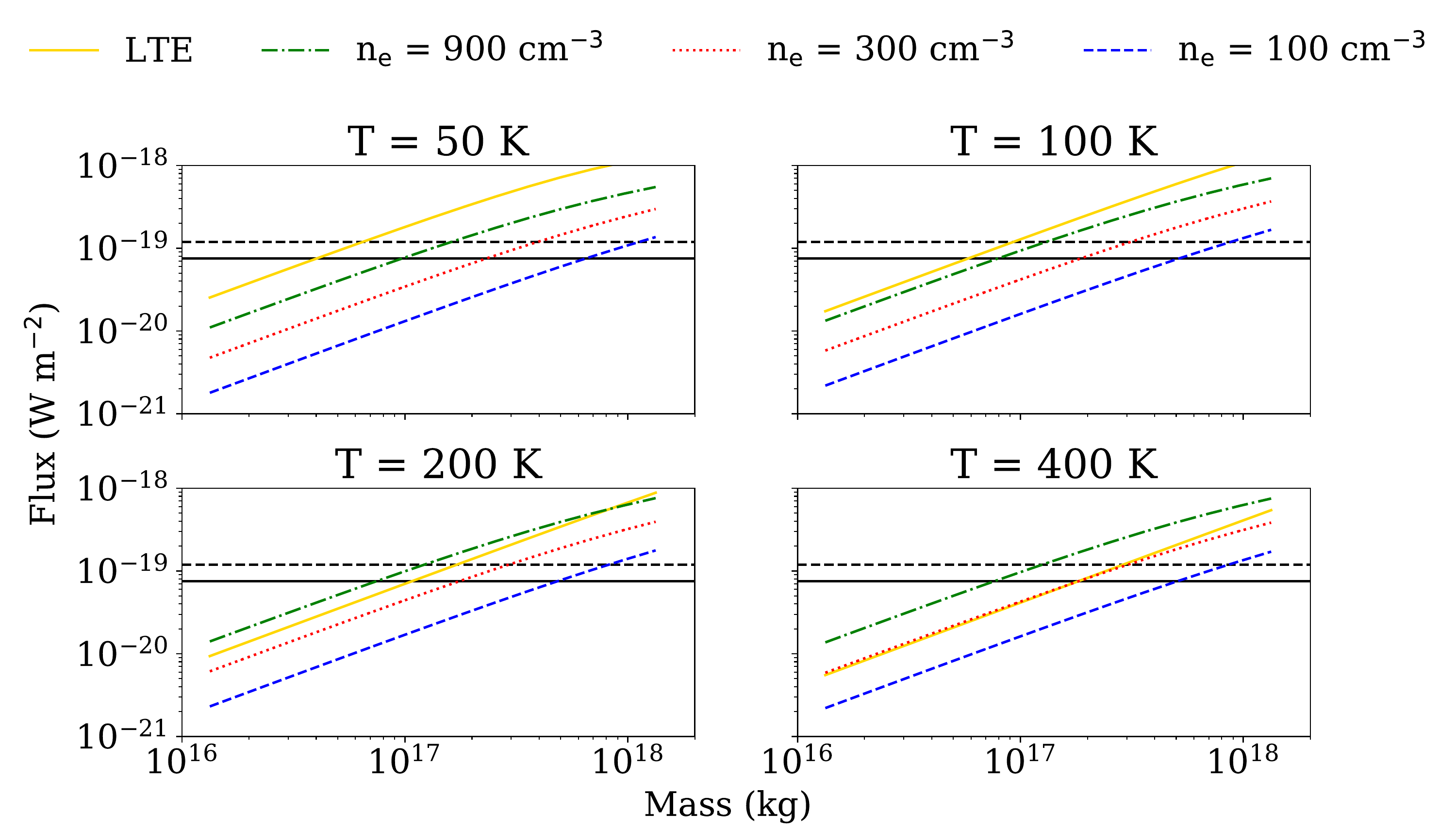}}
	\caption{Computed o-H$_2$O 1$_{10}$--1$_{01}$ flux as a function of H$_2$O gas mass. Four different gas kinetic temperatures
    are shown; 50\,K (top left), 100\,K (top right), 200\,K (bottom left) and 400\,K (bottom right). Solid lines show LTE calculations,
    while each other line corresponds to an electron density; 100\,cm$^{-3}$ (dashed), 300\,cm$^{-3}$ (dotted), and 900\,cm$^{-3}$
    (dash-dotted). The horizontal lines in each plot indicates the 95\% line flux upper limit obtained from the data 
    (Sect.~\ref{subsec:flux_upper_limit}). The solid horizontal line corresponds to using a gaussian profile, and the dashed to the spectral profile matching the CO.  }
	\label{fig:fluxmass}
\end{figure*}

Figure~\ref{fig:fluxmass} shows the calculated flux versus mass, as well as the observed line
flux 95\% upper limits, indicated by the horizontal lines. We take the intersection of
these lines with the curves corresponding to different electron densities as our o-H$_2$O gas mass
upper limits. The measured CO mass of $2.0\times 10^{20}$\,kg \citep{Matra2017}
is then assumed in order to calculate ($T$, $n_{\mathrm{e}}$)-dependent lower limits on the CO/H$_2$O,
as presented in Table~\ref{tab:results}. Finally, Table~\ref{tab:parameters} shows the values of the optical depth at the center of the emission line $\tau_{\nu0}$, the line excitation temperature $T_\mathrm{ex}$, and the molecular column density $N_\mathrm{H_2O}$, all evaluated at the respective gas mas upper limits.

	\begin{table*}
		\caption{H$_2$O gas mass 95\% upper limits and corresponding CO/H$_2$O lower limits.}	
		\label{tab:results}
		\centering
			\begin{tabular}{c c c c c c} 
				\hline \hline
				$T$ & $n_\mathrm{e}$ & Max. H$_2$O, & Min. CO/H$_2$O, & Max. H$_2$O, & Min. CO/H$_2$O,  \\
                (K) & (cm$^{-3}$) & Gaussian profile (kg) & Gaussian profile & CO profile (kg) & CO profile  \\
				\hline
				50	& $100$	& $6.46 \times 10^{17}$ & 310   & $1.13 \times 10^{18}$ & 177   \\			
					& $300$ & $2.32 \times 10^{17}$ & 864   & $3.96 \times 10^{17}$ & 505	\\		
					& $900$ & $9.68 \times 10^{16}$ & 2070  & $1.62 \times 10^{17}$ & 1230  \\
					& LTE	& $4.01 \times 10^{16}$ & 4990  & $6.48 \times 10^{16}$ & 3080  \\
				\cline{1-6	}
				100	& $100$	& $ 5.13 \times 10^{17}$ & 390  & $8.83 \times 10^{17}$ & 226 	\\			
					& $300$ & $ 1.86 \times 10^{17}$ & 1080 & $3.14 \times 10^{17}$ & 638   \\		
					& $900$ & $ 7.91 \times 10^{16}$ & 2530 & $1.31 \times 10^{17}$ & 1530  \\
					& LTE	& $ 5.81 \times 10^{16}$ & 3440 & $9.32 \times 10^{16}$ & 2145  \\
				\cline{1-6	}
				200	& $100$ & $ 4.82 \times 10^{17}$ & 415  & $8.26 \times 10^{17}$ & 242 \\			
					& $300$	& $ 1.75 \times 10^{17}$ & 1140 & $2.94 \times 10^{17}$ & 681   \\		
					& $900$ & $ 7.44 \times 10^{16}$ & 2690 & $1.23 \times 10^{17}$ & 1630  \\
					& LTE   & $ 1.07 \times 10^{17}$ & 1860 & $1.71 \times 10^{17}$ & 1170  \\
				\cline{1-6	}
				400	& $100$	& $ 5.05 \times 10^{17}$ & 396  & $8.64 \times 10^{17}$ & 232   \\			
					& $300$	& $ 1.82 \times 10^{17}$ & 1100 & $3.05 \times 10^{17}$ & 655   \\		
					& $900$ & $ 7.63 \times 10^{16}$ & 2620 & $1.26 \times 10^{17}$ & 1590  \\
					& LTE 	& $ 1.81 \times 10^{17}$ & 1100 & $2.90 \times 10^{17}$ & 690   \\
				\hline
		\end{tabular}
	\end{table*}

\begin{table*}
		\caption{Optical depth $\tau_{\nu0}$, excitation temperature $T_\mathrm{ex}$, and H$_2$O column density $N_{\mathrm{H_2O}}$. Each parameter is evaluated where the expected line flux is equal to the upper limit obtained using the Gaussian or CO spectral profile, respectively. }	
		\label{tab:parameters}
		\centering
			\begin{tabular}{c c c c c c c c} 
				\hline \hline
				$T$ & $n_\mathrm{e}$ & $\tau_{\nu0}$, & $T_\mathrm{ex}$, & $N_{\mathrm{H_2O}}$, & $\tau_{\nu0}$, & $T_\mathrm{ex}$, & $N_{\mathrm{H_2O}}$,  \\
                (K) & (cm$^{-3}$) & Gaussian & Gaussian & Gaussian & CO & CO & CO \\
                & & profile & profile (K) & profile (cm$^{-2}$) & profile & profile (K) & profile (cm$^{-2}$) \\
				\hline
				50	& $100$	& 4.80 & 9.59 & $3.84 \times 10^{13}$ & 7.92 & 11.1  & $6.73 \times 10^{13}$  \\			
					& $300$	& 1.64 & 10.9 & $1.38 \times 10^{13}$ & 2.67 & 12.0 & $2.35 \times 10^{13}$ \\		
					& $900$ & 0.590 & 14.5 & $5.75 \times 10^{12}$ & 0.959 & 15.3 & $9.65 \times 10^{12}$ \\
					& LTE	& $6.56 \times 10^{-2}$ & 50.0 & $2.38 \times 10^{12}$ & 0.106 & 50.0 & $3.85 \times 10^{12}$  \\
				\cline{1-8	}
				100	& $100$	& 3.79 & 9.75 & $ 3.05 \times 10^{13}$ & 6.16 & 11.2 & $5.25 \times 10^{13}$ 	\\			
					& $300$ & 1.29 & 11.5 & $ 1.10 \times 10^{13}$ & 2.07 & 12.5 & $1.86 \times 10^{13}$   \\		
					& $900$ & 0.449 & 16.2 & $ 4.70 \times 10^{12}$ & 0.722 & 16.9 & $7.77 \times 10^{12}$ \\
					& LTE	& $2.80 \times 10^{-2}$ & 100 & $ 3.45 \times 10^{12}$ & $4.5 \times 10^{-2}$ & 100 & $5.54 \times 10^{12}$ \\
				\cline{1-8	}
				200	& $100$	& 3.56 & 9.8 & $ 2.87 \times 10^{13}$ & 5.74  & 11.2 & $4.91 \times 10^{13}$ \\			
					& $300$	& 1.2 & 11.7 & $ 1.04 \times 10^{13}$ & 1.92 & 12.7 & $1.75 \times 10^{13}$  \\		
					& $900$ & 0.411 & 16.8 & $ 4.42 \times 10^{12}$ & 0.659 & 17.5 & $7.23  \times 10^{12}$ \\
					& LTE	& $1.3 \times 10^{-2}$ & 200 & $6.37 \times 10^{12}$ & $2.08 \times 10^{-2}$ & 200 & $1.02 \times 10^{13}$ \\
				\cline{1-8	}
				400	& $100$	& 3.73 & 9.76 & $ 3.00 \times 10^{13}$ & 6.01 & 11.2 & $5.13 \times 10^{13}$ \\			
					& $300$	& 1.25 & 11.6 & $ 1.08 \times 10^{13}$ & 2.00 & 12.6 & $1.81 \times 10^{13}$ \\		
					& $900$ & 0.43 & 16.6 & $ 4.53 \times 10^{12}$ & 0.68 & 17.3 & $7.47 \times 10^{12}$ \\
					& LTE 	& $6.25 \times 10^{-3}$ & 401 & $ 1.08 \times 10^{13}$ & $9.99 \times 10^{-3}$ & 401 & $1.72 \times 10^{13}$ \\
				\hline
		\end{tabular}
	\end{table*}


\section{Discussion}\label{sec:discussion}
A few different trends can be inferred from the plots in Fig.~\ref{fig:fluxmass}. The computed line flux generally increases with both gas mass and electron density, as one would expect. 
For a fixed line flux, the gas mass decreases as the electron density goes up. In the presence of more electrons, the H$_2$O molecules are excited more often, hence fewer of them are needed to account for the flux.

With increasing temperature, other behaviours become apparent. Firstly, the line flux in the LTE case decreases with increasing temperature. This is a result of higher levels becoming increasingly more excited than the ground level, decreasing the emission from the latter. Secondly, all curves flatten out at high masses. This is due to the emission becoming optically thick, hence more mass can be added without raising the flux. The effect is most pronounced for high densities. 

\begin{figure}
	\centering
	\resizebox{\hsize}{!}
	{\includegraphics{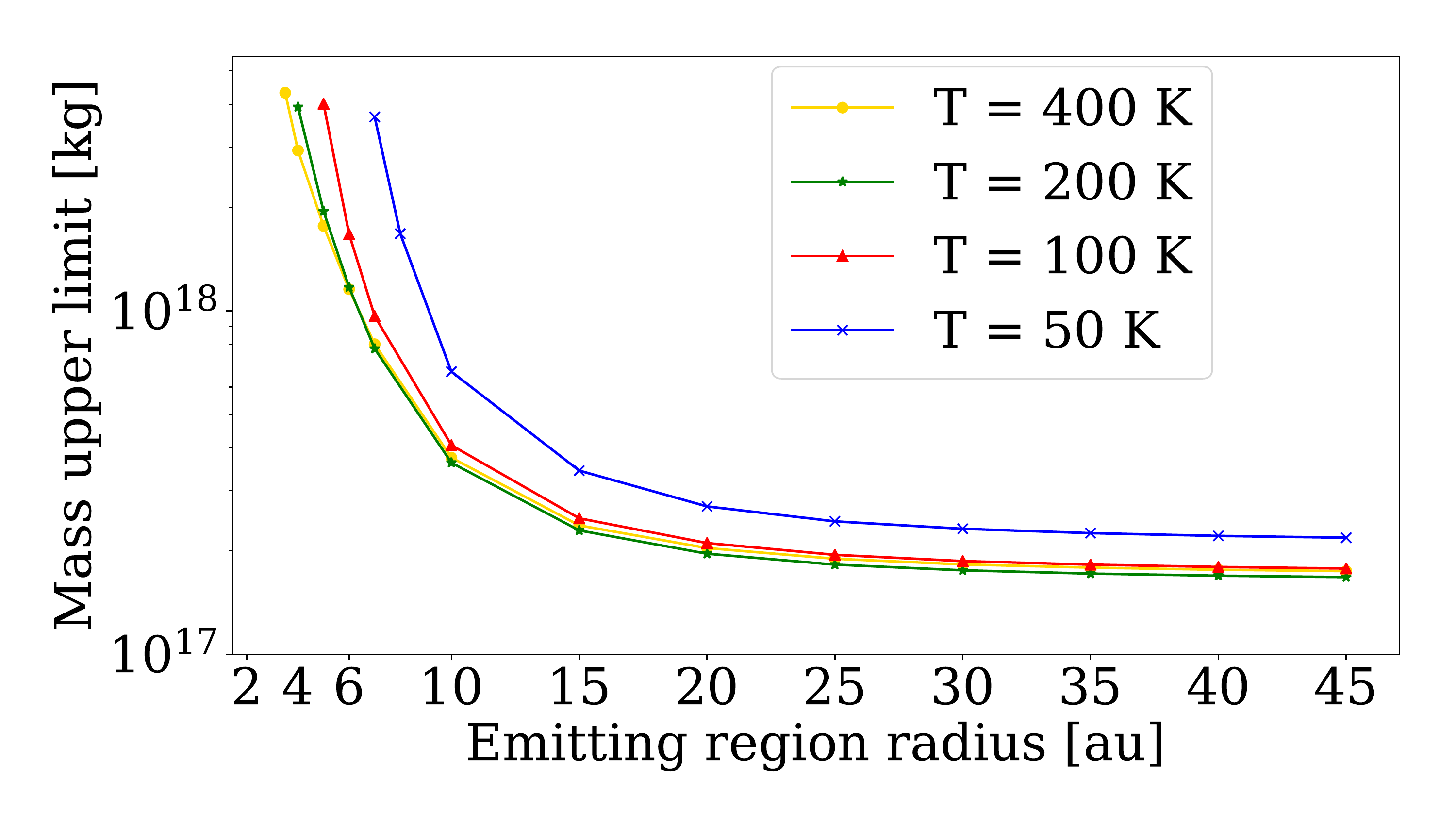}}
	\caption{Derived mass upper limits as a function of emission region radius (assumed to be spherical). In our calculations, we use a radius of 30\,au. The mass uppper limits are for the case of a Gaussian line profile.}
	\label{fig:clumpmass}
\end{figure}

The solid-phase number ratio $\left(N_\mathrm{CO_2}/N_\mathrm{H_2O}\right)_\mathrm{ice}$ in solar system comets has been found to be between 0.002 and 0.23 \citep{Bockelee2017}. Taking into account the relative atomic weights and photodissociation lifetimes at 85\,au in the $\beta$~Pic system of the two species, the gas-phase mass ratio $\left(M_\mathrm{CO_2}/M_\mathrm{H_2O}\right)_\mathrm{gas}$ is determined as:
\begin{equation}
    \left(\frac{M_\mathrm{CO_2}}{M_\mathrm{H_2O}}\right)_\mathrm{gas} = 
    \left(\frac{N_\mathrm{CO_2}}{N_\mathrm{H_2O}}\right)_\mathrm{ice} 
    \frac{\mu_\mathrm{CO}}{\mu_\mathrm{H_2O}}
    \frac{t_\mathrm{CO}}{t_\mathrm{H_2O}} 
     \,\cdot
\end{equation}
This corresponds to gas-phase mass ratios between roughly 16 and 1900, assuming continuous production.

Considering first the case of the narrow spectral line, the observed lower limits are compatible with solar system compositions for all cases of electron densities $n_\mathrm{e}$ = 100, 300\,cm$^{-3}$. ALMA observations of CO lines find $n_\mathrm{e}\sim300$\,cm$^{-3}$ at the location of the CO \citep[Fig.~11 in][]{Matra2017}. None of the cases of $n_\mathrm{e}$ = 900 nor the $T = 50, 100$\,K cases in LTE are compatible with solar system compositions, as the H$_2$O content is too low. 
For the scenario of the broader spectral line, only the two LTE cases with $T = 50, 100$\,K lie outside the expected range from solar system comets. As the gas is not expected to be in LTE, these two cases seem unlikely.
\\
The true distribution of H$_2$O probably lies somewhere between the two extremes we test for. It is difficult to say how likely each scenario is, and also whether the lower limits on the CO/H$_2$O ratio will be pushed upwards by more sensitive observations. Because of the possibility of discrepancies between our derived values and solar system comet values, we here speculate on explanations for this. 

One possible explanation could be episodic gas production; CO has a dissociative lifetime of\,$\sim$50 years, but H$_2$O survives for only 3.5 days. If the gas is produced in intermittent bursts with a frequency lower than once every few days, rather than continuously, we may have caught the system at a time where H$_2$O is gone but CO remains. In that case another observation made just days apart from this one could lead to very different results. Another possibility is that low-velocity collisions between the planetesimals can cause CO to vaporise while leaving the less volatile H$_2$O behind \citep{Dent2014}. However, since the collisions are continually taking place, the planetesimals will eventually be ground down to dust grains, from which the H$_2$O is efficiently UV photo-sputtered by $\beta$~Pic \citep{Grigorieva2007}. A third possibility is that the clump radius is significantly smaller than our assumed 30\,au, in which case the increased optical depth would mean that more mass can be present without violating the upper limit on the flux, particularly so for low temperatures. We plot the inferred mass upper limit, for the case of a narrow spectral line, against emission region radius in Fig.~\ref{fig:clumpmass}. We see that the radius would have to shrink to less than about 15\,au for the derived mass to change significantly.
Finally, the non-detection of H$_2$O could be caused by a different initial composition of the colliders compared to solar system comets. The $\beta$~Pic system has several times been observed to have peculiar gas abundances. Carbon is particularly overabundant compared to Fe \citep{Roberge2006,Brandeker2011, Cataldi2018}, and oxygen is likely overabundant as well \citep{Brandeker2016}.
Integrating the spatial number density of Fe from \citet{Nilsson2012}, we find the total Fe mass to be $M_{\mathrm{Fe}} = 3.0\times10^{19}$\,kg. Comparing to
a possible mass range of C \citep[$M_{\mathrm{C}} = \lbrack 3.0-18.5 \rbrack\times 10^{21}$\,kg,][]{Cataldi2018},
a C/Fe number ratio of 50--330 times solar values is obtained. Given the different spatial distributions found for C and Fe, perhaps this is just a reflection
of different origins, where the C could be mostly produced by dissociation of CO and CO$_2$. If so, the O abundance would likewise mostly be formed by 
dissociation of CO, CO$_2$, and H$_2$O, and an estimate of the O abundance could be used to indirectly constrain the H$_2$O/CO ratio.
Unfortunately, the O mass is difficult to derive since the detected \ion{O}{i} 63\,$\mu$m emission is optically thick and strongly dependent on the
spatial distribution and excitation conditions. Other dissociation products that could be searched for are H and OH, both of which are accessible for
sensitive measurements at radio frequencies by the planned \textit{Square Kilometer Array}.


\section{Summary}\label{sec:conclusions}
We present a HIFI non-detection of H$_2$O emission in the $\beta$~Pictoris system. We find the 95\% upper limit on the o-H$_2$O 1$_{10}$--1$_{01}$ line flux to
be $7.5 \times 10^{-20}$\,W\,m$^{-2}$ or $1.2 \times 10^{-19}$\,W\,m$^{-2}$, depending on the assumed production region of H$_2$O. We test for two scenarios: one where all H$_2$O production occurs at the place of the previously observed CO clump, and one where the production follows the entire CO spatial profile.

Calculating the level populations for the case of an electron density of $300\,\mathrm{cm}^{-3}$ this is translated to an upper limit on the H$_2$O gas mass of $(1.75$--$3.96) \times 10^{17} \, \mathrm{kg}$, depending on both the production region and the kinetic gas temperature, which has previously been constrained to be between 40 and a few hundred K \citep{Matra2017}. 

From the mass upper limits we derive lower limits on the CO/H$_2$O mass ratio. Assuming again $n_\mathrm{e}\,\sim300$\,cm$^{-3}$, the ratios in the scenario of a clump production region fall between 864 and 1140. For the spread-out production scenario, the ratios are between 505 and 681. Since the emission is thought to originate from colliding comet-like bodies, we can compare these lower limits to known solar system values of 16--1900 \citep{Bockelee2017}. In the case of this chosen electron density, the ratios thus fall within the solar system composition range. For electron densities of 900\,cm$^{-3}$ and the cases of LTE with $T=50,100$\,K , all lower limits on CO/H$_2$O in the clump production scenario are larger than expected from solar system values, indicating water-poor conditions. In the spread-out production scenario, this is only true for an emission region in LTE with temperatures of $T = 50,100$\,K. If the comets of $\beta$~Pic are indeed water-poor, possible explanations are episodic gas production or simply an initial composition different from solar system comets.

Since the now decommissioned HIFI instrument was unique in targeting the H$_2$O 1$_{10}$--1$_{01}$ line, more sensitive measurements will have to await future instrumentation. 
Measuring the dissociation products OH, H, or O could potentially indirectly constrain the water content of the colliding comets. The future \textit{Square Kilometer Array} will be uniquely sensitive to both H and OH.

\begin{acknowledgements}
We thank Alexandre~Faure for providing collisional excitation rates at low temperatures, and Luca Matrà for sharing the model of the dust radiation field and the ALMA data of CO. We also thank the referee for helpful and thorough comments. HIFI has been designed and built by a consortium of institutes and university departments from across Europe, Canada and the United States under the leadership of SRON Netherlands Institute for Space Research, Groningen, The Netherlands and with major contributions from Germany, France and the US. Consortium members are: Canada: CSA, U.Waterloo; France: CESR, LAB, LERMA, IRAM; Germany: KOSMA, MPIfR, MPS; Ireland, NUI Maynooth; Italy: ASI, IFSI-INAF, Osservatorio Astrofisico di Arcetri-INAF; Netherlands: SRON, TUD; Poland: CAMK, CBK; Spain: Observatorio Astronómico Nacional (IGN), Centro de Astrobiología (CSIC-INTA). Sweden: Chalmers University of Technology - MC2, RSS \& GARD; Onsala Space Observatory; Swedish National Space Board, Stockholm University - Stockholm Observatory; Switzerland: ETH Zurich, FHNW; USA: Caltech, JPL, NHSC. This work was supported by the Momentum grant of the MTA CSFK Lend\"ulet Disk Research Group and by JSPS KAKENHI Grant Number JP16F16770. This research has made use of NASA's Astrophysics Data System.
\end{acknowledgements}

\bibliographystyle{aa}
\bibliography{Cavallius_2019_bib}
\end{document}